\documentclass[doublecol]{epl2} 

\usepackage{amsmath}
\usepackage{float}

\title{Time-dependent condensation of bosonic dysprosium}
\shorttitle{Time-dependent condensation of bosonic dysprosium} 

\author{Max Regalado Kloos \and Georg Wolschin}
\shortauthor{M. Regalado Kloos, G. Wolschin}

\institute{                   
Institute for Theoretical Physics, Heidelberg University, Philosophenweg 16, Heidelberg, 69120, Germany
}

\abstract
{We investigate thermalization and time-dependent Bose--Einstein condensate formation in ultracold $^{164}$Dy using a nonlinear boson diffusion equation. As compared to alkali atoms such as $^{39}$K or $^{87}$Rb, the strong magnetic dipole interaction modifies the scattering-length dependence of the transport coefficients that govern thermalization and condensate formation. A prediction for the time-dependent condensate fraction in $^{164}$Dy is made.}

\begin{document}

\maketitle

\section{Introduction}
\label{intro}
Early theoretical models \cite{sto91,gz97} for time-dependent Bose--Einstein condensate (BEC) formation in alkali atoms expected the process to be  governed by the elastic cross section for $s$-wave scattering. It has turned out, however, that
the scattering length $a$ rather than $a^2$ is the decisive quantity because of the emerging coherence between the highly occupied infrared (IR) modes \cite{dwg17}. This has recently been underlined very clearly by time-dependent deep-quench BEC formation 
experiments for various scattering lengths with bosonic potassium atoms in a box trap, and tunable interactions \cite{gli21}. 

The buildup of coherence can generally not be accounted for explicitly in Boltzmann-like kinetic approaches \cite{setk95} that do not keep track of the phases of the atomic wave functions. It has, however,  been possible to model the measured time dependence of condensate formation in alkali atoms based on a nonlinear boson diffusion equation (NBDE) \cite{gw18,kgw22,lgw24}
with empirical diffusion and drift coefficients
that are taken to be proportional to the respective $s$-wave scattering lengths rather than the cross sections. In these works, we have 
computed the time-dependent condensate fraction from the condition that the total number of particles in the nonequilibrium cloud and in the condensate must be conserved in time following the quench.

Whereas the role of the magnetic moment in the thermalization and condensate-formation process is negligible for ultracold alkali atoms, 
the situation is different in strongly dipolar quantum gases of lanthanides such as $^{164}$Dy \cite{lu11} with a magnetic dipole moment of $d\simeq9.93\,\mu_\mathrm{B}$ and a corresponding
dipolar length $a_\mathrm{dd}\simeq 130.8\,a_0$.
Here, the isotropic contact interaction is supplemented by the long-range anisotropic dipole-dipole interaction, changing the expected time-dependence of BEC formation for different scattering lengths. 
So far, no time-dependent data for BEC formation are available for such a system, but based on the experience with alkali atoms, we present in this Letter a prediction for 
thermalization and condensate formation of $^{164}$Dy in a parabolic trap that can later be compared with data. 

We use the nonlinear diffusion model with constant coefficients for a schematic prediction. In this limit, the NBDE can be solved exactly through a nonlinear transformation for sufficiently simple initial conditions \cite{gw18,gw20,rgw20} -- once data become
available, this can be supplemented by a numerical model with energy-dependent coefficients. To provide exact solutions in the constant-coefficient case, we model a quench with a 
truncated Bose--Einstein initial occupation-number distribution and density of states in a parabolic trap as initial condition. The presently available condensation experiments with dipolar atomic Bose-Einstein condensates (dBEC) such as dysprosium or erbium \cite{lu11,ka16,fe16,cho16,luc18} use gradual evaporative cooling that is more difficult to model using the NBDE, and no time-dependent 
data are available yet.
\section{The model}
\label{model}
Our nonequilibrium-statistical description of thermalization and time-dependent condensate formation in ultracold atomic vapours is based on a nonlinear partial differential equation that has been derived earlier \cite{gw18}.  Starting from the many-body level, the problem is reduced to the one-body level, with a mean-field Hamiltonian plus confining potential that represents the trap. A two-body collision term causes thermalization, and also drives the time-dependent formation of the Bose--Einstein condensate once the system is cooled below the critical temperature, such that quantum coherence sets in. 

The occupation-number distribution is obtained from the diagonal elements of the ensemble-averaged one-body density operator, and it can be further simplified through the so-called ergodic approximation  \cite{snowo89,svi91,lrw96,jgz97}, such that the distribution functions depend only on energy, and time. 
In this limit, the collision term still correctly accounts for bosonic enhancement in the infrared region through its nonlinear contribution. The time evolution of the single-particle 
distribution functions $n\equiv \langle n(\epsilon,t)\rangle$ is given by the solution of the nonlinear boson diffusion equation \cite{gw18}
 \begin{equation}
\partial_t {n}_{t}=-\partial_\epsilon\bigl[v\,n\,(1+ n)+n\,\partial_\epsilon D\bigr]+\partial_{\epsilon\epsilon}\bigl[D\,n\bigr]\,.
 \label{boseq}
\end{equation}
The coefficient $D\,(\epsilon,t)$ accounts for diffusion of particles in the energy space. It is related to the drift  $v\,(\epsilon,t)$
through the fluctuation-dissipation relation ($\hbar=c=k_\mathrm{B}=1$)
 \begin{equation}
 \label{fdr}
\lim_{t\rightarrow \infty}[-v\,(\epsilon,t)/D\,(\epsilon,t)] \equiv 1/T_\mathrm{f}\,,
\end{equation}
where $T_\mathrm{f}$ is the final temperature reached at equilibrium.
The transport coefficients contain the many-body physics and generally depend on energy and time. When derived from a quantum Boltzmann equation, they will depend on the second moment of the interaction \cite{gw18}, but do not account explicitly for the buildup of coherence in the system. However, coherence can be phenomenologically introduced by taking $D,v$ as being proportional to the interaction strengths ($s$-wave scattering lengths $a$) as previously proposed \cite{kgw22,lgw24}, rather than $\propto a^2$.

The NBDE in the above form accounts for the time-dependent 
thermalization in the cloud with particle number $N_\mathrm{th}(t)$. We shall consider condensate formation via elastic collisions  indirectly through particle-number conservation at each timestep,
 \begin{equation}
N_\mathrm{c}(t) = N_\mathrm{i} – N_\mathrm{th}(t)\,,
\end{equation}
with the initial particle number $N_\mathrm{i}$ that remains in the system following a quench.
The effect of interactions between particle cloud and condensate is implicit through the choice of the mesoscopic 
transport coefficients $D, v.$

It has been shown analytically \cite{lgw24} that the stationary solution $n_\infty(\epsilon)$ of the nonlinear diffusion equation that is attained for $t\rightarrow\infty$ equals the Bose--Einstein equilibrium distribution $n_\mathrm{eq}(\epsilon)$ 
\begin{equation}
n_\infty(\epsilon)=n_\mathrm{eq}(\epsilon)=\frac{1}{e^{(\epsilon-\mu)/T}- 1}\,,
 \label{Bose--Einstein}
\end{equation}
provided the ratio $v/D$ has no energy dependence, as required by the above fluctuation-dissipation relation, Eq.\,(\ref{fdr}).
The chemical potential is $\mu\leq0$ in a finite Bose system, it appears as an integration constant in the solution of 
Eq.\,(\ref{boseq}) for $t\rightarrow\infty$. 
For an energy-dependent diffusion coefficient, the term $n\,\partial D/\partial \epsilon$  in Eq.\,(\ref{boseq}) 
ensures the correct equilibrium limit \cite{lgw24}.

Even in the simplified case of constant transport coefficients, the NBDE
preserves  the essential features of Bose--Einstein statistics that are contained in the bosonic quantum Boltzmann equation 
during the time evolution. 
Moreover, Eq.\,(\ref{boseq}) is one of the few nonlinear partial differential equations with a clear physical meaning that can be solved exactly through a nonlinear transformation \cite{gw18,rgw20}. The solution can be written as
\begin{eqnarray}
        n(\epsilon,t) = T\, \partial_\epsilon\ln{\mathcal{Z}(\epsilon,t)} - \frac{1}{2}=  \frac{T}{\mathcal{Z}}\, \partial_\epsilon \mathcal{Z} - \frac{1}{2}\,.
    \label{net} 
    \end{eqnarray}
The time-dependent partition function ${\mathcal{Z}(\epsilon,t)}$  obeys a linear diffusion equation $\partial_t\mathcal{Z}(\epsilon,t)=D\partial_{\epsilon\epsilon}\mathcal{Z}(\epsilon,t)$, which has a Gaussian Green's function $G(\epsilon,x,t)$. The resulting partition function is an integral over the Green's function and an exponential function $F(x)$ that contains an energy integral over the initial conditions \cite{gw18,rgw20}. For bosons, one has to consider the boundary conditions at the singularity $\epsilon = \mu \leq 0$ \cite{gw20} when solving the linear diffusion equation as a combined initial- and boundary-value problem. The boundary conditions can be expressed as
 \(\lim_{\epsilon \searrow \mu} n(\epsilon,t) = \infty\) \,$\forall$ \(t\). One obtains a vanishing partition function at the boundary \( \mathcal{Z} (\mu,t) = 0\), and the energy range is restricted to  $\epsilon \ge \mu$. For sufficiently simple initial conditions like a truncated Bose--Einstein distribution, the integral over the initial conditions, the calculations of ${\mathcal{Z}(\epsilon,t)}$  and its energy derivative, and thus, of the occupation-number distributions can be carried out exactly.
\section{Thermalization of ultracold atoms}
Regarding the thermalization of ultracold atoms via number-conserving elastic collisions, the initial nonequilibrium distribution following a sudden quench at $\epsilon=\epsilon_\mathrm{i}$ can be represented as a Bose--Einstein equilibrium distribution with an energy cutoff, $n(\epsilon,0)=n_\infty(\epsilon)\,\theta(\epsilon-\epsilon_\mathrm{i})$. Here we make use of  the exact solution of the nonlinear boson diffusion equation for the combined initial- and boundary value problem as derived previously \cite{rgw20}, with the time-dependent partition function written as an infinite series
\begin{eqnarray}
{\mathcal{Z}}(\epsilon,t)&=&\sqrt{4 D t} \, \exp\Bigl(-\frac{\mu}{2 T_\mathrm{f}}\Bigr)\qquad\qquad\\ \nonumber
 &\times&\sum_{k=0}^{\infty} \dbinom{\frac{T_\mathrm{i}}{T_\mathrm{f}}}{k} (-1)^k \times
f_k^{{T}_\mathrm{i},{T}_\mathrm{f}} \,(\epsilon,t)\,.
\label{partfct}
\end{eqnarray}
The analytical expressions for $f_k^{{T}_\mathrm{i},{T}_\mathrm{f}} \,(\epsilon,t)$ 
are combinations of exponentials and error functions. A similar result exists for the derivative of ${\mathcal{Z}}(\epsilon,t)$ \cite{rgw20}. The series terminates at $k_\mathrm{max}$ if $T_\mathrm{i}/T_\mathrm{f}$ is an integer, as in thermalization and condensate formation of ultracold $^{39}$K atoms \cite{gli21,kgw22,lgw24}, where $k_\mathrm{max}=T_\mathrm{i}/T_\mathrm{f}=130\,\mathrm{nK}\,/32.5\,\mathrm{nK}=4$. 

For a deep quench in ultracold alkali atoms {with negligible dipolar interactions}, we had previously calculated the time-dependent single-particle distribution functions based on this analytical approach.  Using energy-independent  transport coefficients for various scattering lengths $a$ of the contact interaction, analytical results for the single-particle distribution functions $n(\epsilon,t)$ in $^{39}$K were presented in Ref.\,\cite{kgw22} (see Figs.\,1/2 there). Numerical solutions of the NBDE with energy-dependent coefficients have been calculated in Ref.\,\cite{lgw24}. 

Together with the condition that the overall particle number in the cloud and in the condensate is conserved in time following the quench, the approach allows us to account for the measured \cite{gli21} time- and scattering-length-dependent condensate fraction up to and beyond the point when statistical equilibrium is reached: For different interaction strengths as tuned by an external magnetic field in the experiment with alkali atoms, we have taken the transport coefficients $D,v$ to be proportional to the contact-interaction scattering lengths $a$ (rather than the cross sections) to account for the emerging coherence, thus achieving consistency with the data for the time-dependent condensate fractions. 

The equilibration times are then proportional to $1/a$ because $\tau_\mathrm{eq}\propto D/v^2$, consistent with condensate-fraction data for alkali atoms.
We now take this agreement as a justification to apply our method for a prediction in dipolar atoms such as $^{164}$Dy, where no data are available yet for the time-dependent condensate fraction.
\section{Ultracold dipolar dysprosium}
\label{udy}
 In dipolar atoms, the thermalization of the atoms in the cloud following evaporative cooling has been measured e.g. for $^{162,164}$Dy \cite{ta16} and for the most abundant bosonic atoms of erbium \cite{pa22}. However, the time dependence of an emerging condensate for temperatures below the critical value has not yet been observed for lanthanide atoms with a strong dipolar interaction. 

In atoms such as $^{164}$Dy with a large dipole magnetic moment $\mu\simeq9.93\,\mu_\mathrm{B}$, there is an interplay between the contact interaction that is decisive for thermalization in alkali atoms, and the magnetic dipole interaction. The scattering cross section of aligned dipolar atoms or molecules in a low-temperature gas becomes independent of energy \cite{bo09}, and it can be expressed in the Born approximation apart from the contribution due to the $s$-wave scattering length. This is a universal behaviour for all dipolar atoms and molecules.

The total interaction potential can be written for aligned dipoles as in case of an external magnetic field
\begin{align}
U(\textbf{r})&=U_\mathrm{contact}(\textbf{r})+U_\mathrm{dd}(\textbf{r})\\ \nonumber
&=\frac{4\pi a_s}{m}\delta(\textbf{r})+\frac{\mu_0\mu_\mathrm{Dy}^2}{4\pi}\frac{(1-3\cos^2\theta)}{|\textbf{r}|^3}
\end{align}
with the $s$-wave scattering length $a_s$ of the contact interaction, magnetic permeability $\mu_0$, magnetic dipole moment $\mu_\mathrm{Dy}$, angle $\theta$ and distance $r=|\textbf{r}|$ between the centers of two dipoles that are aligned in the external magnetic field. Given the very large de Broglie wavelength of the atoms at ultracold temperatures,  higher partial waves are suppressed due to the centrifugal barrier.

With the dipolar length for magnetic dipole-dipole interaction of $^{164}$Dy
\begin{equation}
a_\mathrm{dd}=\mu_0\mu_\mathrm{Dy}^2 m_\mathrm{Dy}/(12\pi)\simeq 130.8\,a_0\,,
\end{equation}
the elastic cross section of bosonic $^{164}$Dy atoms can be obtained in Born approximation \cite{bo09,bj14,sb15}. The perturbative approximation is applicable because the strong dipole-dipole interaction becomes weak at long range due to its proportionality to $\propto 1/r^3$ -- unlike long-range Coulomb-like $1/r$ interactions, where the perturbative approximation would not be valid. However, there is no barrier in $s$-wave scattering and hence, the contribution of the contact interaction with the scattering length $a_s$  is added separately. Averaging the dipole-dipole interaction over all incident angles, the momentum-independent result for the elastic cross section is given by the sum of the squared moduli of both scattering lengths with prefactors
\cite{bj14,sb15}
\begin{equation}
\sigma_\mathrm{el}=8\pi a_s^2+\frac{32\pi}{45}\left(\frac{3}{2}a_\mathrm{dd}\right)^2\,,
\end{equation}
where 
the dipolar part includes the sum of even and odd partial-wave contributions \cite{bj14}.

As in our previous phenomenological consideration of coherence effects during time-dependent condensate formation in case of alkali atoms where $D,v\propto \sqrt{\sigma_\mathrm{el}}\propto a_s$  \cite{kgw22,lgw24}, one should expect that the transport coefficients in the dysprosium case are $D,v\propto \sqrt{\sigma_\mathrm{el}}$, and accordingly, the time scales  $\tau_\mathrm{ini}, \tau_\mathrm{eq}\propto 1/\sqrt{\sigma_\mathrm{el}}$, and we shall base our present predictions for $^{164}$Dy on this proposition.
 \begin{figure}
\begin{center}
 \includegraphics[scale=0.26]{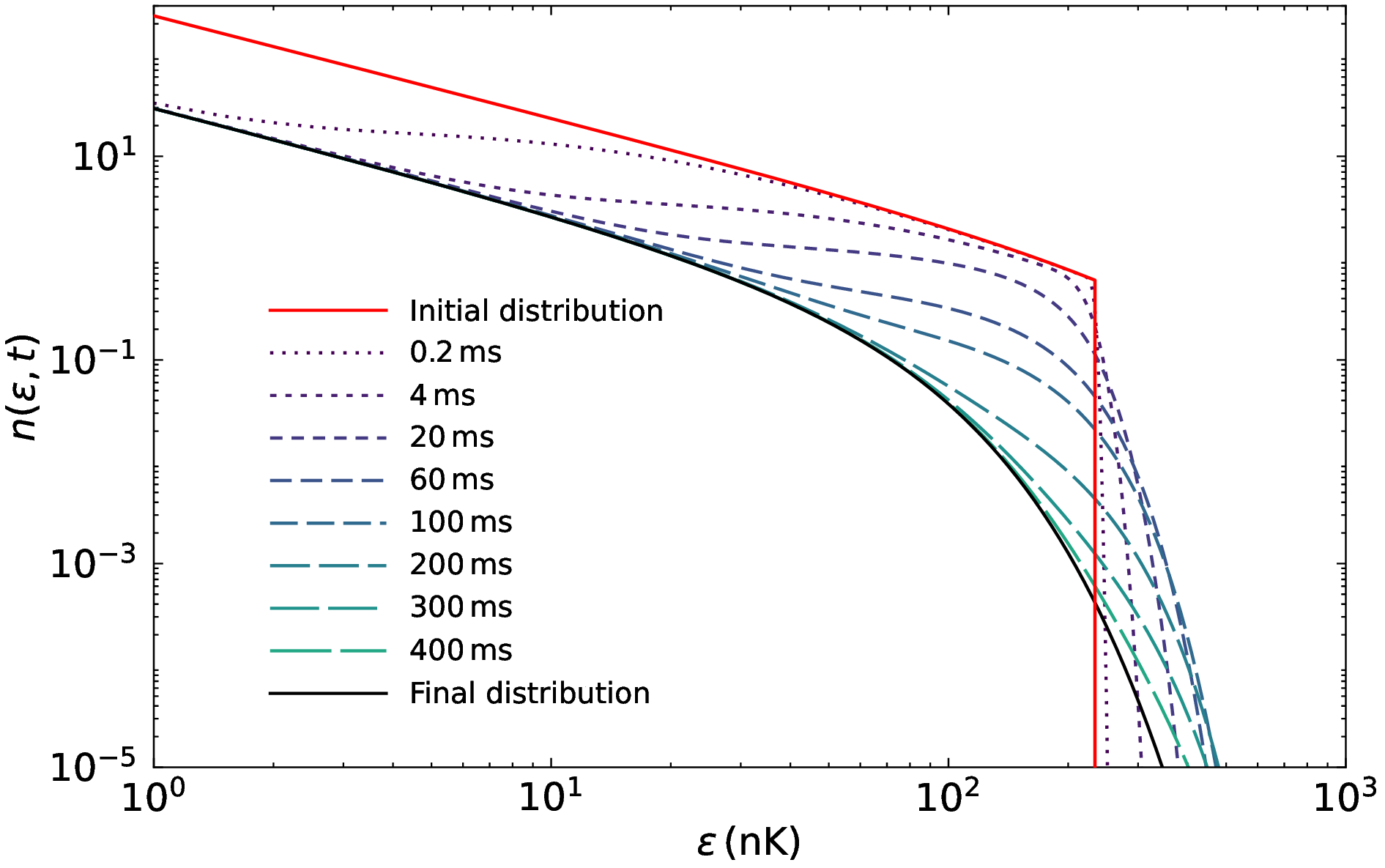}%
\caption{\label{fig1} {{Thermalization of $^{164}$Dy.} Nonequilibrium evolution of a quenched $^{164}$Dy gas in double-log scale for eight timesteps (increasing dash lengths) as calculated from analytical NBDE solutions. The system is taken to evolve from an initial truncated Bose--Einstein distribution with $T_\mathrm{i}=240$ nK, $\epsilon_\mathrm{i}=234$ nK (here for $\mu_\mathrm{i}=0$) to the equilibrium distribution with $T_\mathrm{i}=30$ nK and $\mu=0$. The transport coefficients are  $D=45$ nK$^2$/ms, $v=-1.5$ nK/ms, the equilibration time at the cut is $\tau_\mathrm{eq}\simeq600$ ms for a scattering 
parameter $\sqrt{\sigma_\mathrm{el}}/a_0=637.90$.}}
\end{center}
\end{figure}
 \begin{figure}
 \begin{center}
 \includegraphics[scale=0.235]{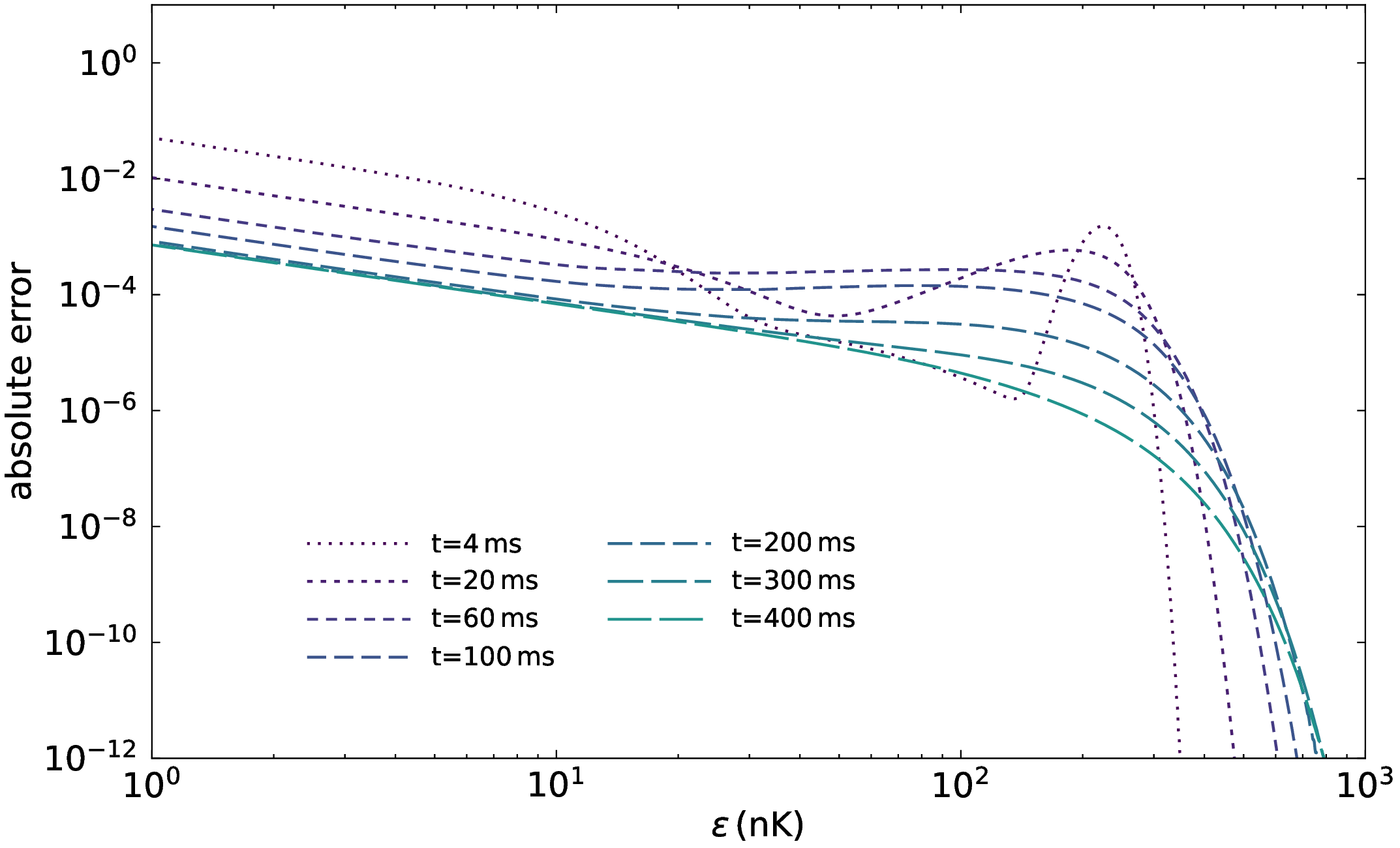}%
\caption{\label{fig2} {Analytical \textit{vs.} numerical NBDE solution for $^{164}$Dy.} The absolute errors of the numerical NBDE solutions relative to the analytical solutions shown in Fig.\,\ref{fig1} are displayed for different times. They are defined as the absolute difference between the analytical
and the numerical solution $|n_\mathrm{ana}(\epsilon,t)-n_\mathrm{num}(\epsilon,t)|.$}
\end{center}
\end{figure}

First we solve the NBDE analytically to obtain time-dependent single-particle distribution functions $n(\epsilon,t)$ in $^{164}$Dy vapour for a quench in a parabolic trap with density of states $g(\epsilon)\propto \epsilon^2$. Given an external magnetic field of 2.52 G close to a
  Feshbach resonance, the $s$-wave scattering length is $a_s\simeq113\,a_0$. Adding the contribution from the dipole-dipole interaction, we obtain $a_\mathrm{tot}\equiv\sqrt{\sigma_\mathrm{el}}=637.90\,a_0$. We calculate the occupation-number distributions for $T_\mathrm{i}=240$ nK, $T_\mathrm{f}=30$ nK, $\mu_\mathrm{i}=-753.56$ nK, $\epsilon_\mathrm{i}=234$ nK (hence, $\eta=\epsilon_\mathrm{i}/T_\mathrm{i}=0.975$), $D=45$\,nK$^2$/ms, $v=-1.5$\,nK/ms. 
  
    Results are shown in Fig.\,\ref{fig1} for eight timesteps between the initial truncated distribution, and the final Bose--Einstein equilibrium result, which is smoothly approached in the time evolution. The equilibration time at the cut $\epsilon=\epsilon_\mathrm{i}=234$ nK is $\tau_\mathrm{eq}\simeq 600$ ms. It is seen that thermalization in the infrared  is much faster than in the ultraviolet (UV) energy region, where the analytical solutions for $t\gg \tau_\mathrm{eq}$ still differ from the equilibrium limit. As already discussed \cite{lgw24} for alkali atoms, this is largely a consequence of our assumption of constant transport coefficients and can be modified with energy-dependent coefficients that cause faster thermalization in the UV.

Energy-dependent transport coefficients require, however, to numerically solve the NBDE. For constant transport coefficients, we have already tested the accuracy of numerical solutions with a corresponding newly written Julia program against the exact results, see Fig.\,\ref{fig2}.
The numerical results reproduce the analytical ones rather precisely, but -- in particular at short times -- with rising inaccuracies towards the singularity and at the quench. Here, we proceed with the constant-coefficient case and its exact solutions as shown in Fig.\,\ref{fig1} for the occupation-number distributions $n(\epsilon,t)$ to now calculate the time-dependent condensate fraction. 
\section{Time-dependent condensate fraction}
\label{tcf}
 \begin{figure}
\centering
 \includegraphics[scale=0.26]{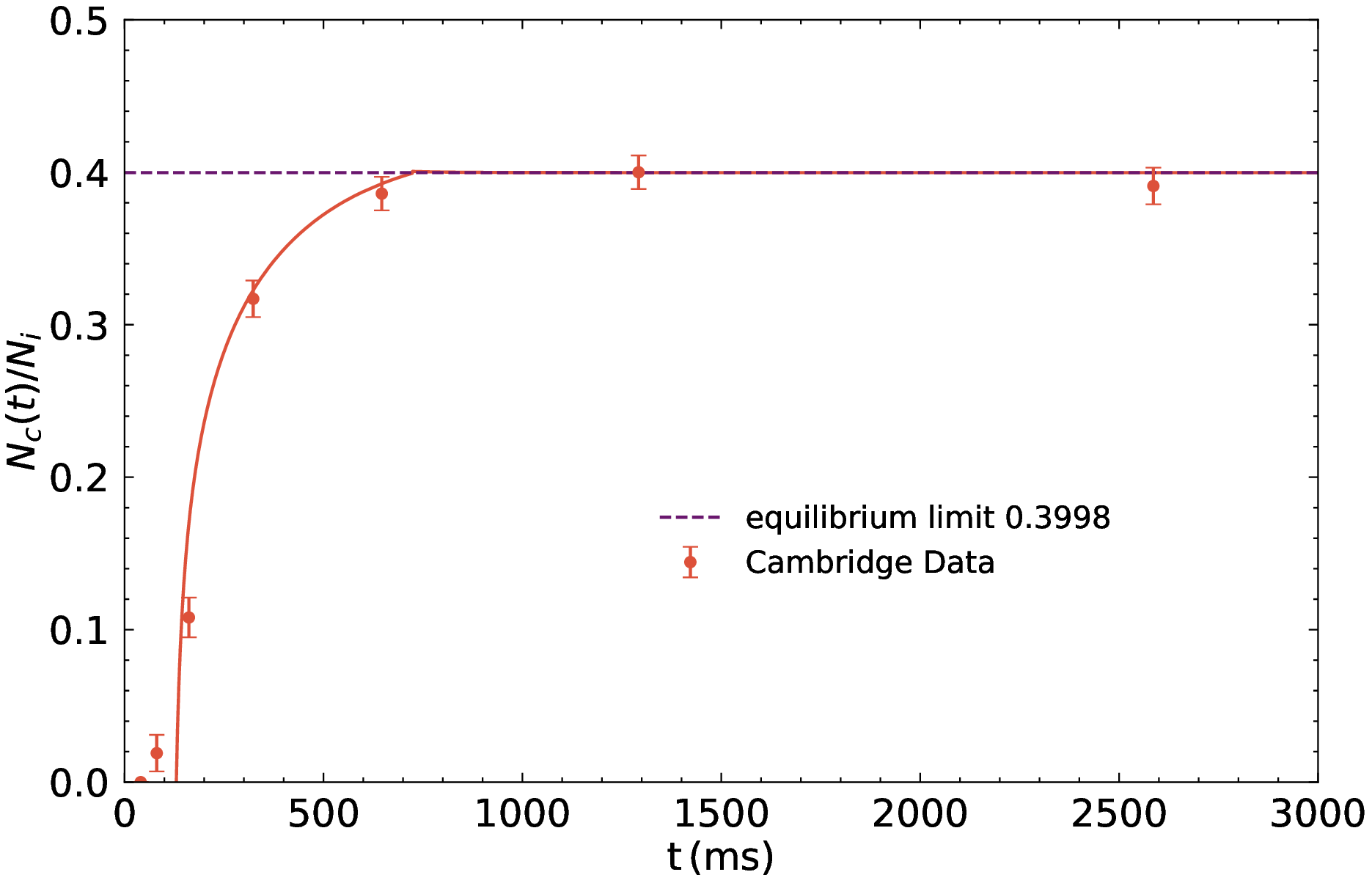}%
\caption{\label{fig3} {Time-dependent condensate fraction for an equilibrating 3D $^{39}$K Bose gas.}  The calculated condensate fraction based on the analytical solutions of the NBDE (see text) is displayed as function of time (solid curve) and compared to Cambridge data \cite{gli21}  for a scattering length $a/a_0=140$ (symbols, with error bars reflecting experimental fitting errors.). In the experiment, ultracold $^{39}$K atoms with temperature $T_\mathrm{i} = 130$ nK in a box trap are subjected to a rapid quench, subsequently thermalize and  -- starting at $\tau_\mathrm{ini}$=130\,ms -- form a BEC with $\simeq 40\%$ particle content in the statistical equilibrium limit ($T_\mathrm{f}=32.5$ nK).}
\end{figure}
 \begin{figure}
 \centering
 \includegraphics[scale=0.278]{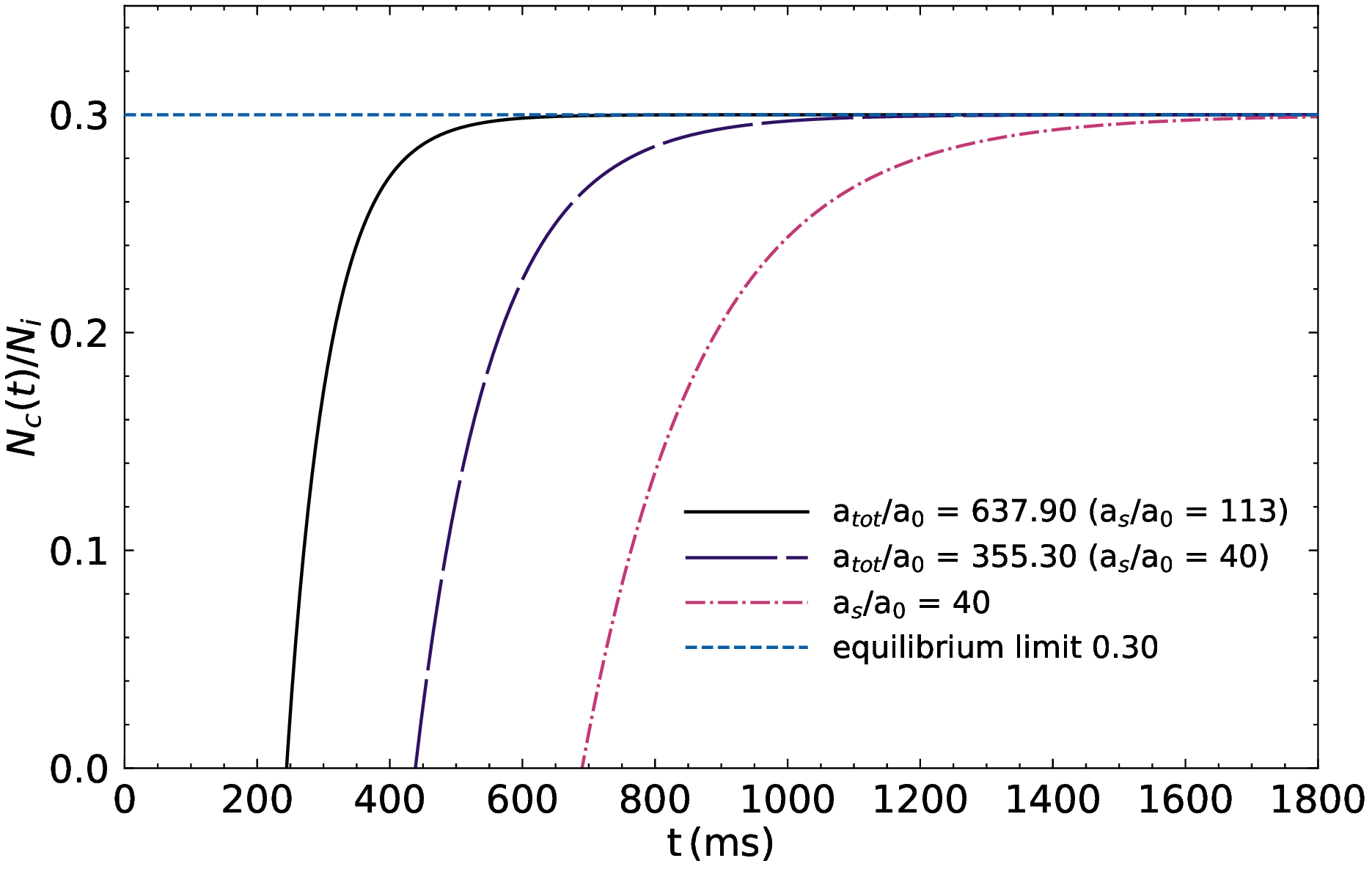}%
\caption{\label{fig4} {Time-dependent condensate fraction for an equilibrating 3D $^{164}$Dy Bose gas.}  The calculations are based on the analytical solutions of the NBDE (see text) and displayed as functions of time for  $a_\mathrm{tot}/a_0\equiv\sqrt{\sigma_\mathrm{el}}/a_0\simeq 637.90$, (solid curve, {with $a_s/a_0=113$}). Transport coefficients have then been {recalculated for}  a reduced contact-interaction scattering length {$a_\mathrm{s}/a_0=40$ (dot-dashed curve), and the correspondingly reduced total scattering length $a_\mathrm{tot}/a_0=355.30$}  (dashed curve). {The significant difference in scaling of the time-dependent condensate fraction with $a_s$ \textit{vs.} $a_\mathrm{tot}$ is expected to be measurable. It indicates the coherence mechanism (see text).}}
\end{figure}
The conserved total particle number $N_\mathrm{i}$ following the quench consists of the time-dependent particle numbers $N_\mathrm{th}(t)$ in the cloud, and  $N_\mathrm{c}(t)$ in the condensate. Hence, we calculate 
 the time-dependent condensate fraction as
 \begin{equation}
  N_\mathrm{c}(t)/N_\mathrm{i}=1-N_\mathrm{th}(t)/N_\mathrm{i}=1-\int_0^\infty g(\epsilon)\,{n}(\epsilon,t)\,{d}\epsilon/N_\mathrm{i}\,,
  \label{cf}
  \end{equation}
  where $N_\mathrm{i}=\int_0^\infty g(\epsilon)\,{n}(\epsilon,t=0)\,{d}\epsilon$ is the initial particle number just after the quench, and $N_\mathrm{th}(t)$ the integral of the time-dependent single-particle distribution functions $n(\epsilon,t)$  once the chemical potential has reached the value $\mu=0$ such that condensate formation starts. 
 
  The equilibrium value of the condensate fraction is determined by the initial particle number $N_\mathrm{i}$ and the final temperature $T_\mathrm{f}$ as
\begin{equation}
N^\mathrm{eq}_\mathrm{c}/N_\mathrm{i}=1-g_0T_\mathrm{f}^{3/2}\zeta(3/2)\Gamma(3/2)/N_\mathrm{i}\,.
\label{cfe}
\end{equation}

In a parabolic trap as taken in our present calculation for $^{164}$Dy, the density of states in 3D is $g(\epsilon)\propto\epsilon^2$, whereas in
a three-dimensional cylindrical box trap, $g(\epsilon)=g_0\sqrt{\epsilon}$.
In the latter case, time-dependent condensate-formation data for $^{39}$K \cite{gli21} can be compared with our model. The result is shown in Fig.\,\ref{fig3} for a scattering length of $a_s=140\,a_0$ as in Ref.\,\cite{kgw22}, but now up to very large times
 $t=3000$ ms $\gg\tau_\mathrm{eq}=600$ ms where equilibrium persists.

In this calculation, the solutions of the NBDE for constant coefficients would show an
overshoot of the time-dependent
condensate fraction above the equilibrium value $N^\mathrm{eq}_\mathrm{c}/N_\mathrm{i}$, before reaching that value from above. This
 is a consequence of the fact that the time-dependent solution of the NBDE for constant coefficients equilibrates too slowly in the UV. To avoid the overshoot, we calculate the distribution function from the exact NBDE solution ${n}(\epsilon,t)$ up to $\epsilon=\epsilon_c$ with $\epsilon_c=27.5$ nK for $^{39}$K and $t=\tau_\mathrm{eq}$, but use an approach to the equilibrium distribution according to a  linear relaxation ansatz with time constant $\tau_\mathrm{rel}=90$ ms for $\epsilon>\epsilon_c$ and $t>\tau_\mathrm{eq}$. 
This is in line with the expectation that bosonic enhancement and condensate formation -- which are implicit in the nonlinear diffusion equation -- become less relevant at higher energies, such that a relaxation ansatz for the time evolution with a short relaxation time $\tau_\mathrm{rel}$ can be taken. 

With this modification, the condensate fraction is found to be in agreement with the data, Fig.\,\ref{fig3}. In a parabolic trap with a steeper energy dependence of the density of states $\propto \epsilon^2$ as in our present $^{164}$Dy calculation, no such modification for constant transport coefficients is required to avoid an overshoot of the condensate fraction above the equilibrium value.
The agreement with the $^{39}$K data confirms, in particular, the proportionality of the transport coefficients to the contact-interaction scattering lengths $a_s$, and the dependence of the time scales on $1/a_s$, rather than $1/a_s^2$. This indirect proof of emerging coherence in the formation of the condensate in alkali atoms has already been presented in Fig.\,3 of Ref.\,\cite{kgw22}. 

 Calculating the time-dependent condensate fraction for $^{164}$Dy as before in case of the alkali atoms, we now take the dipole-dipole interaction into account together with the contact-interaction strengths $a_s/a_0=113$, because both are expected to contribute to the thermalization and condensate formation. Averaging over the incident angles results in $a_\mathrm{tot}/a_0\equiv\sqrt{\sigma_\mathrm{el}}/a_0\simeq 637.90$. We thus obtain the result shown in Fig.\,\ref{fig4}, solid curve. As in Fig.\,\ref{fig1}, the equilibration time is $\tau_\mathrm{eq}\simeq 600$ ms. The initiation time is $\tau_\mathrm{ini}\simeq 240$ ms.

Our treatment of dipolar interactions amounts to a replacement of the contact $s$-wave scattering length with the square root of the elastic cross section that includes a dipolar contribution. In case comparison with forthcoming data will confirm this, there would be no qualitative change in 3D thermalization from dipolar interactions, but instead just a renormalization of the relevant length and time scales. 
 
To determine whether the onset of coherence is solely mediated by the contact interaction, or also by the medium-range magnetic dipole interaction, we propose
to investigate the scaling of condensate formation when the $s$-wave scattering length $a_\mathrm{s}$ is modified. As a specific example, we {reduce the contact-interaction scattering length from $a_s/a_0=113$ to $a_s/a_0=40$}, which could in principle be reached by tuning the external magnetic field. 

In the dot-dashed curve in Fig.\,\ref{fig4}, we have accordingly scaled the transport coefficients {$D, v$} with {the corresponding ratio $40/113$}, and the times with {$\tau\propto D/v^2=113/40$}. We thus assume that  coherence is mediated solely by the short-range contact interaction. If scaling was determined based on the elastic cross section {rather than the scattering length} as was proposed in some of the early theories for BEC formation, the characteristic times for condensate formation (relative to the ones for $a_\mathrm{s}/a_0=113$) would be too large to fit into the range shown in Fig.\,\ref{fig4}. 
 
In case the medium-range dipolar interaction contributes not only to thermalization, but also to the buildup of coherence, one expects, however, a scaling with the effective scattering length $a_\mathrm{tot}/a_0\equiv\sqrt{\sigma_\mathrm{el}}/a_0$ {when the $s$-wave scattering length is reduced to $a_s/a_0=40$ by tuning the external magnetic field}, dashed curve in Fig.\,\ref{fig4}. {Since the difference in scaling is significant enough to be measurable, future experiments could decide the issue}\footnote{For an increase of $a_s/a_0$ above 113, the effect is not sufficiently pronounced to be measurable}. This will determine whether the buildup of coherence in time-dependent dipolar atomic condensate formation occurs only in the $s$-wave channel as for {atoms such as $^{39}$K that have negligible dipolar interactions} \cite{gli21,ma25},
or is also mediated by the strong magnetic dipole interaction. 
 
 Our results should thus be taken as predictions, to be compared with forthcoming time-dependent condensate-formation data in dipolar atoms as functions of $a_\mathrm{tot}$ or $a_\mathrm{s}$ for different contact-interaction scattering lengths, as tuned by an external magnetic field. There is a clear and measurable difference between our predictions for a scaling with contact-interaction scattering lengths only, \textit{vs.} scaling with the effective total scattering lengths including the dipolar interaction, and experiments should decide between these two limiting possibilities. The difference determines whether coherence in time-dependent condensate formation is being built up only through the contact interaction, or mediated as well by the medium-range magnetic dipole interaction that includes higher partial waves.

It would also be of interest to investigate the 2D problem, with dipoles aligned in \textit{vs.} out-of-plane. In this case it is, however,  not possible to average over the incident angles, as done here for the 3D situation, such that the present model must be extended accordingly. \\\\
\acknowledgments

MRK thanks Julian Kusch for the collaboration during his MSc time, and Christian G{\"o}lzh{\"a}user for information about the current Heidelberg dysprosium experiment. GW acknowledges discussions with Lauriane Chomaz about BEC formation in ultracold dysprosium, and with Pascal Naidon at RIKEN about the scattering-length dependence of the transport coefficients.

\bibliographystyle{eplbib}
\bibliography{gw_25.bib}


\end{document}